\documentclass[10pt,twocolumn,superscriptaddress,showpacs,amsmath,amssymb,osajnl,floatfix]{revtex4-1}

\usepackage{epsfig}
\usepackage{amsmath}
\usepackage{amssymb}
\usepackage{bm}
\usepackage{graphicx}
\usepackage{color}
\usepackage{graphicx,rotating}
\usepackage{txfonts}
\usepackage{microtype}
\usepackage{ulem}

\begin{document}

\title{Gamma-ray beams with large orbital angular momentum via nonlinear Compton \\scattering with radiation reaction}

\author{Yue-Yue Chen}
\email{yue-yue.chen@mpi-hd.mpg.de}
\affiliation{Max-Planck-Institut f\"{u}r Kernphysik, Saupfercheckweg 1, 69117 Heidelberg, Germany}
\author{Jian-Xing Li}
\affiliation{Max-Planck-Institut f\"{u}r Kernphysik, Saupfercheckweg 1, 69117 Heidelberg, Germany}
\affiliation{School of Science, Xi'an Jiaotong University, Xi'an 710049, China}
\author{Karen Z. Hatsagortsyan}
\email{k.hatsagortsyan@mpi-hd.mpg.de}
\affiliation{Max-Planck-Institut f\"{u}r Kernphysik, Saupfercheckweg 1, 69117 Heidelberg, Germany}
\author{Christoph H. Keitel}
\affiliation{Max-Planck-Institut f\"{u}r Kernphysik, Saupfercheckweg 1, 69117 Heidelberg, Germany}

\begin{abstract}

Gamma-ray beams with large angular momentum are a very valuable tool to study astrophysical phenomena in a laboratory. We investigate  generation of well-collimated $\gamma$-ray beams with a very large orbital angular momentum using nonlinear Compton scattering of a strong laser pulse of twisted photons at ultra-relativistic electrons. Angular momentum conservation among absorbed laser photons, quantum radiation and electrons are numerically demonstrated in the quantum radiation dominated regime. We point out that the angular momentum of the absorbed laser photons is not solely transferred to the emitted $\gamma$-photons, but due to radiation reaction shared between the $\gamma$-photons and interacting electrons. The efficiency of the angular momentum transfer is optimized with respect to the laser and  electron beam parameters. The accompanying process of electron-positron pair production is furthermore shown to enhance the orbital angular momentum gained by the $\gamma$-ray beam.

\end{abstract}

\maketitle

Vortex light is an electromagnetic field that carries orbital angular momentum (OAM). It has a spiral phase ramp around a singularity, and a Poynting vector resembling a corkscrew, rotating about the propagation axis \cite{allen1992orbital}. The OAM of optical vortices can be transferred to atoms, molecules, and nanostructures \cite{1367-2630-12-8-083053, alexandrescu2006mechanism,babiker2002orbital,toyoda2013transfer,Surzhykov2015,Muller2016,Afanasev2013,Schmiegelow2016,Peshkov2017}, which allows diverse applications in the visible and infrared regime, such as in quantum information \cite{mair2001entanglement}, microscopy \cite{jesacher2005shadow}, micromanipulation \cite{he1995direct} and in the detection of spinning objects \cite{lavery2013detection}. To bring these applications down to the nanometer and atomic scale \cite{van2007prediction}, different methods are proposed to generate extreme-ultraviolet and x-ray vortex beams via high-order harmonic generation in a gas-phase atomic target \cite{Wang2015,hernandez2013attosecond,gauthier2017tunable}, plasma \cite{zhang2015generation,Zhang_2016,Denoeud2017}, and helical undulator \cite{sasaki2008proposal,hemsing2013,ribivc2017extreme}. 
Recently, very promising concepts were developed to upgrade near-IR vortex beams to very high intensities \cite{sueda2004laguerre,lin2016generation,vieira2016amplification}, which provides the possibility to generate $\gamma$-ray vortex beams using Compton/Thomoson scattering of twisted light off ultra-relativistic electrons
\cite{jentschura2011generation,jentschura2011compton,ivanov2011scattering,Stock2015,
petrillo2016compton,taira2017gamma}.

Twisted $\gamma$-photons once realized could have straightforward applications to drive specific nuclear transitions \cite{taira2017gamma}, and to control and manipulate the rotation of nuclear matter \cite{jentschura2011compton}.
In another prospect, $\gamma$-ray beams with ultrahigh OAM could impact the dynamics of rotating astrophysical objects, such as rotating black holes and luminous pulsars.  This kind of $\gamma$-ray beams can be generated in an astrophysical environment \cite{tamburini2011twisting,hartemann2000high,
stewart1972non,arons1972nonlinear,takiwaki2009special}. For example, the radiation from an accretion disk around a Kerr black hole experiences a well-defined phase variation and polarization rotation due to gravitational effects and therefore has both spin and orbital angular momentum \cite{tamburini2011twisting}. When these twisted photons travel through the corona region, they may experience inverse Compton scattering with the fast-moving particles giving rise to gamma photons. Moreover, radiation emitted by luminous pulsars and quasars propagate through inhomogeneous surroundings, experiencing behavior analogous to light propagating through a spiral phase plate or a holographic phase plate \cite{harwit2003photon}. The acquired OAM of radiation can be transfered to high-energy emission by nonlinear inverse Compton scattering \cite{arons1972nonlinear}. For a simulation of this type of phenomena in laboratory astrophysics the first step is to produce $\gamma$-ray beams with a large OAM by Compton scattering off twisted light, which is the topic of our investigation.

\begin{figure}[b]
	\begin{center}
	\includegraphics[width=0.5\textwidth]{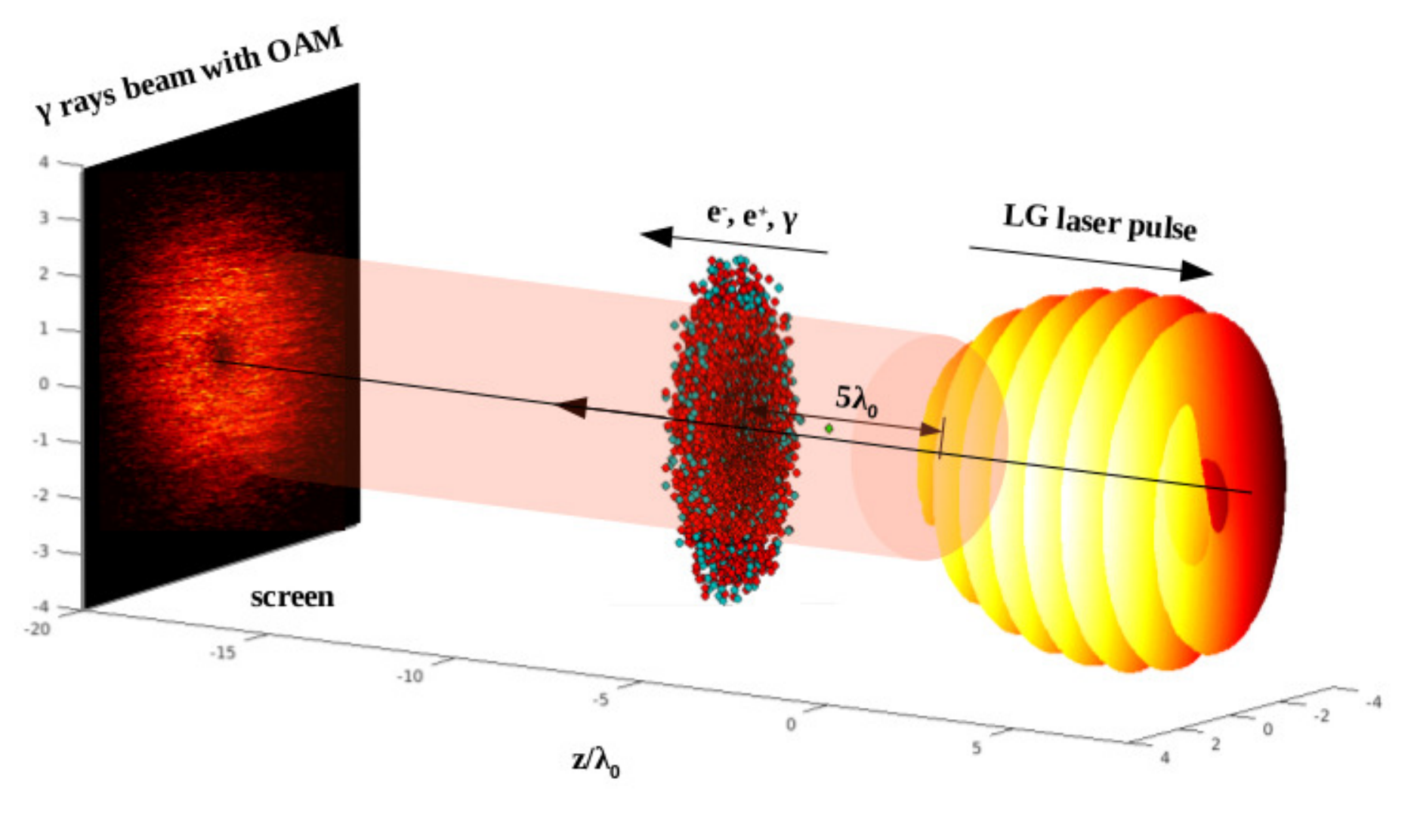}
	\end{center}
	\caption{Scheme for generation of a $\gamma$-ray beam with OAM. An intense Laguerre-Gaussian laser beam of  linear [illustrated in the figure] or circular polarization counterpropagates with and scatters off an electron beam. Electrons absorb multiple laser photons generating a $\gamma$-photon.  The spin and OAM of laser photons are transferred to the electrons and in this way to the  $\gamma$-photon beam. The radiation energy distribution  is illustrated on the screen.}
	\label{scheme}
\end{figure}

The emitted photons due to Compton scattering of the vortex laser beam  by an electron will be twisted, if during the photon formation  the electron experiences the  vortex structure of the laser field. This will be the case if the electron can be represented as a plane wave, or when the photon formation length is comparable with the laser wavelength.  The latter takes place when the laser field parameter is not large, $\xi\equiv eE_0/(mc\omega_0)\lesssim 1$ \cite{RMP_2012}, where $E_0$ and $\omega_0$ are the laser field amplitude and the frequency, respectively, $-e$ and $m$ are the electron charge and mass, respectively, and $c$ is the speed of light. The perturbative regime of Compton scattering, $\xi\ll 1$, when one laser twisted photon is scattered off an ultra-relativistic electron into a twisted $\gamma$-photon, with a  topological charge similar to the incoming laser field is considered in \cite{jentschura2011generation,jentschura2011compton,ivanov2011scattering}.
The topological charge of the emitted twisted $\gamma$-photons can be increased using multiphoton phenomena in  stronger laser fields. In the classical regime x- and $\gamma$-rays with OAM have been demonstrated by Thomoson scattering  of laser pulses with either orbital or spin angular momentum (SAM) at a moderate nonlinearity with $\xi\sim 1$ \cite{petrillo2016compton,taira2017gamma}. In stronger laser fields with $\xi\gg 1$  \cite{vieira2016amplification,danson2015petawatt} not only the number of scattered photons, but also the nonlinearity is dramatically increased, which may enhance significantly the OAM of the emitted photon beam. Using the interaction of intense twisted lasers with plasma, $\gamma$-rays with a large OAM are envisaged \cite{liu2016generation,vieira2016high,gong2017ultra}. However, the twisted $\gamma$-photon emission is not highly energetic in {\color{blue} \cite{liu2016generation,vieira2016high},} and not collimated in \cite{gong2017ultra}.

In the ultraintense regime $\xi\gg 1$, the coherence length of the photon emission in nonlinear Compton process is $\xi$ time less than the laser period. As a consequence the electron during the emission of the $\gamma$-photon experiences the laser field as an almost constant field, rather than the structure of the laser field which carries the information on the OAM. Therefore, the emitted single $\gamma$-photon is not twisted, i.e., not in a certain angular momentum state. However, the beam of  incoherent $\gamma$-photons produced via the laser scattering by a beam of electrons possess OAM with respect to the propagation axis.

\begin{figure}
	\begin{center}
	\includegraphics[width=0.5\textwidth]{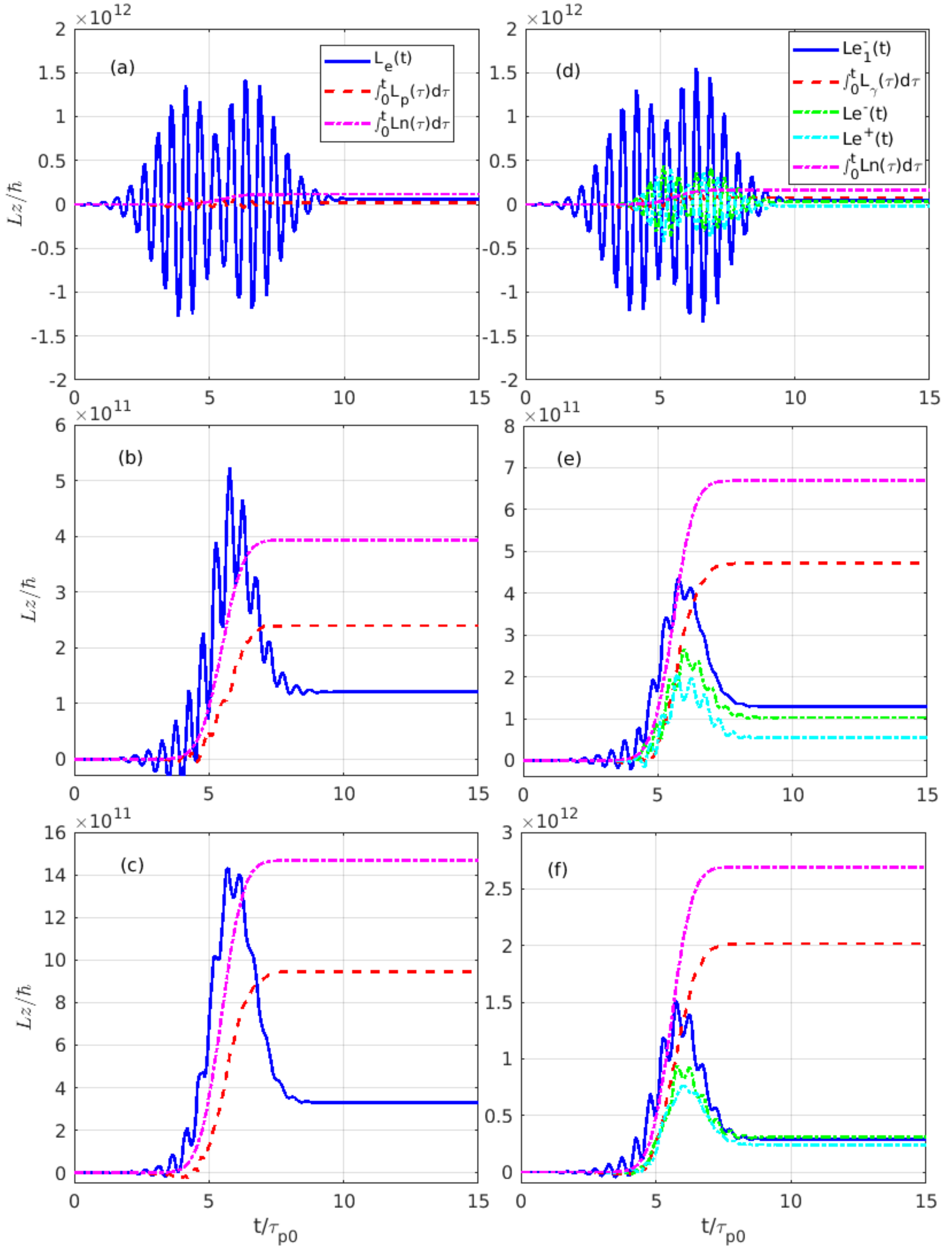}
	\end{center}
	\caption{Angular momentum of $\gamma$-photons (red-dashed), electrons (blue-solid), and the absorbed laser photons (magenta-dash-dotted) in the  LG$_{01}$ laser field of linear (top), and circular (bottom) polarization, and in a circularly polarized Gaussian laser field (middle), with accounting for the pair production (right column) and without (left column). The OAM of the created electrons and positrons are shown by (green-dash-dotted) and (light-blue-dash-dotted), respectively. The laser and electron parameters are $\xi=120, \gamma=10^4$.}
	\label{Fig. 2}
\end{figure}

In this letter, we investigate Compton scattering of an intense laser beam of twisted photons by ultra-relativistic electrons in the quantum radiation dominated regime (QRDR), see Fig.~\ref{scheme}, i.e., when the radiation energy during a laser period is comparable with the electron energy. For this purpose, the angular momentum evolution of the quantum radiation is calculated numerically with a semi-classical method, where the electron dynamics in the laser field is calculated classically, and the photon emission and possible further pair production via quantum electrodynamics with the Monte Carlo algorithm in \cite{Elkina_2011,Ridgers_2014,Green_2015}. As a key result, a high energy $\gamma$-photon beam with both very high OAM and collimation is generated. In contrast to previous works \cite{jentschura2011generation,jentschura2011compton,ivanov2011scattering,Stock2015,petrillo2016compton,taira2017gamma} we further demonstrate that when radiation reaction is accounted for, part of the OAM and SAM of the absorbed laser photons is transferred to the electron beam. Moreover, the accompanying pair production process is shown to cause a counterintuitive increase of the OAM of the $\gamma$-beam due to extra absorption of twisted laser photons by secondary particles.

We consider interaction of an intense Laguerre-Gaussian (LG) laser pulse  (circularly or linearly polarized) with a counterpropagating ultrarelativistic electron beam. The interaction is in the QRDR  \cite{RMP_2012}, when $\alpha\xi\chi\gtrsim 1$  and the quantum nonlinearity parameter $\chi\gtrsim 1$ [the emitted photon recoil and the further pair production are not negligible], with fine structure constant $\alpha$, $\chi\approx 2(\omega_0/m)\xi\gamma$, and the electron Lorentz factor $\gamma$. 
The generated $\gamma$-ray beam with OAM is well collimated in the regime $\gamma \gg \xi$, with the emission angle $\theta\sim \xi/\gamma$ \cite{RMP_2012}.

The field of LG mode in the paraxial approximation reads \cite{allen1992orbital,allen1996spin}(see also \cite{april2008nonparaxial}):
  \begin{eqnarray}
   E_{p\ell}(r,\phi,z)&=&i\omega_0\left[(\alpha\hat{\textbf{x}}+
 \beta\hat{\textbf{y}})u_{p\ell}-\frac{i}{k_0}\left(\alpha\frac{\partial u_{p\ell}}{\partial x}+\beta\frac{\partial u_{p\ell}}{\partial y}\right)\hat{\textbf{z}}\right]\\
 &\times&\exp\left[i(\omega_0 t-k_0z)\right]\exp\left[-\left(\frac{2\sqrt{\ln2}(\omega_0 t-k_0z)}{\omega_0\tau_p}\right)^2\right],\nonumber
\end{eqnarray}
where $k_0=2\pi/\lambda_0=\omega_0/c$ is the wave-vector, $\tau_p$  the pulse duration, $\sigma_z\equiv i(\alpha\beta^*-\alpha^*\beta)$  [$\sigma_z=0$  for linear, and $\mp 1$ for right/left-hand circular polarization], and 
 \begin{eqnarray}
  u_{p\ell}(r,\phi,z)&=&\frac{C}{(1+z^2/z_R^2)^{1/2}}\left(\frac{r\sqrt{2}}{w(z)}\right)^\ell L_p^\ell\left(\frac{2r^2}{w^2(z)}\right)\nonumber \\
 &\times&\exp\left[-\frac{r^2}{w^2(z)}\right]\exp\left[\frac{-ikr^2z}{2(z^2+z_R^2)}\right]\exp(-i\ell\phi)\\
 &\times&\exp\left[i(2p+\ell+1)\tan^{-1}\left(\frac{z}{z_R}\right)\right],\nonumber
  \end{eqnarray}
with $w(z)=\sqrt{1+z^2/z_R^2}$, the Rayleigh length $z_R=\pi w_0^2/\lambda_0^2$ and the laser waist size $w_0$.
Each photon of a LG$_{p\ell}$ beam carries $\hbar\sigma_z$ of SAM and $\hbar \ell$ of OAM \cite{allen1992orbital}, where $\ell$ is the topological charge. The laser parameters are $\xi=120$ and $\gamma=10^4$, $\lambda_0=800$~nm, $\chi\approx 4$, $\ell=1$, $w_0=2\lambda_0$, and $\omega_0\tau_p/2\pi=6$. 
The  electron beam, with a length of $\lambda_0$ and a radius of $4\lambda_0$, consists of 2$\times 10^5$ electrons and has a transverse spatial Gaussian distribution with a width of $\sigma_{\bot}=1.2\lambda_0$.

The total emission energy in the case of a linearly polarized LG mode is shown in Fig. \ref{scheme}.
The energy distribution of the $\gamma$-ray beam has a ring-shaped intensity profile, indicating that it carries OAM. The quantitative evaluation of OAM is presented in Fig.~\ref{Fig. 2} for three cases: a LG laser pulse with linear and circular polarization, and a circularly polarized Gaussian laser field with $\sigma_z=1$. The OAM of $\gamma$-photons,  electrons and positrons with respect to the $z$-axis are calculated with $L_z=xp_y-yp_x$, where $p_x$ and $p_y$ are the components of the linear momenta, and $x,y$ are the coordinates of the particles (for the $\gamma$-photon it is the emission coordinate, as the formation length of the photon is well localized in this ultrarelativistic regime). The total angular momentum absorbed from the laser is $L_n=(\ell+\sigma_z)n\hbar$, where $n$ is the number of the absorbed laser photons. The latter is calculated from the energy-momentum conservation $q+n k=q'+k'$, where $q$, $q'$, and $k$ are the 4-quasi-momenta of incoming and outgoing electrons, and the emitted $\gamma$-photon, respectively. The electron's quasi-momenta are estimated via their relationship to the instantaneous electron  momentum, see \cite{Supplement}.

\begin{table}[b]
\centering
\caption{ OAM [in units of $\hbar$] and energy changes [in units of $mc^2$] of electrons and photons after the interaction,  according to Fig. \ref{Fig. 2}. Left and right subcollunms correspond to without and with pair production, respectively. Subscripts $e$ and $\gamma$ denote charged particles and emitted photons, respectively, $n$ is the number of absorbed laser photons due to the $\gamma$-photon emission,  $\tilde{n}\equiv \Delta E_{e}/\hbar\omega_0$ is the  energy change described by a photon number,  $N_{\gamma}$ is 
the $\gamma$-photon number, $\overline{\ell}$ is the average topological number.} 
\label{Tab. 1}
\begin{tabular}{c|c|c|c|c}
\hline
\hline
 & \multicolumn{2}{c|}{circular Gaussian} & \multicolumn{2}{c}{circular LG$_{01}$} \\\hline
$ L_{e}+L_{\gamma}$ & $3.61\times10^{11}$ & $7.57\times10^{11}$ & $1.28\times10^{12}$ & $2.86\times10^{12}$ \\
$\int_{0}^{t}L_\gamma(\tau)d\tau$ & $2.4\times10^{11}$ & $4.72\times10^{11}$ & $9.46\times10^{11}$ & $2.02\times10^{12}$ \\
$\Delta E_{e}+E_{\gamma}$ & $5.31\times10^{5}$ & $1.06\times10^{6}$ & $1.02\times10^{6}$ & $2.25\times10^{6}$ \\
$\int_{0}^{t}E_\gamma(\tau)d\tau$ & $1.65\times10^{9}$ & $1.54\times10^{9}$ & $1.8\times10^{9}$ & $1.67\times10^{9}$ \\
$n$ & $3.93\times10^{11}$ & $6.69\times10^{11}$ & $7.33\times10^{11}$ & $1.35\times10^{12}$ \\
 $\tilde{n} $ & $2.19\times10^{11}$ & $4.38\times10^{11}$ & $4.21\times10^{11}$ & $9.29\times10^{11}$\\
$N_{\gamma}$ & $3.51\times10^{6}$ & $4.86\times10^{6}$ & $4.29\times10^{6}$ & $6.69\times10^{6}$ \\
$\overline{\ell}$ & $6.74\times10^{4}$ & $9.6\times10^{4}$ & $2.19\times10^{5}$ & $2.99\times10^{5}$ 
\\\hline
\hline
\end{tabular}
\end{table}
\begin{figure}
	\begin{center}
	\includegraphics[width=0.5\textwidth]{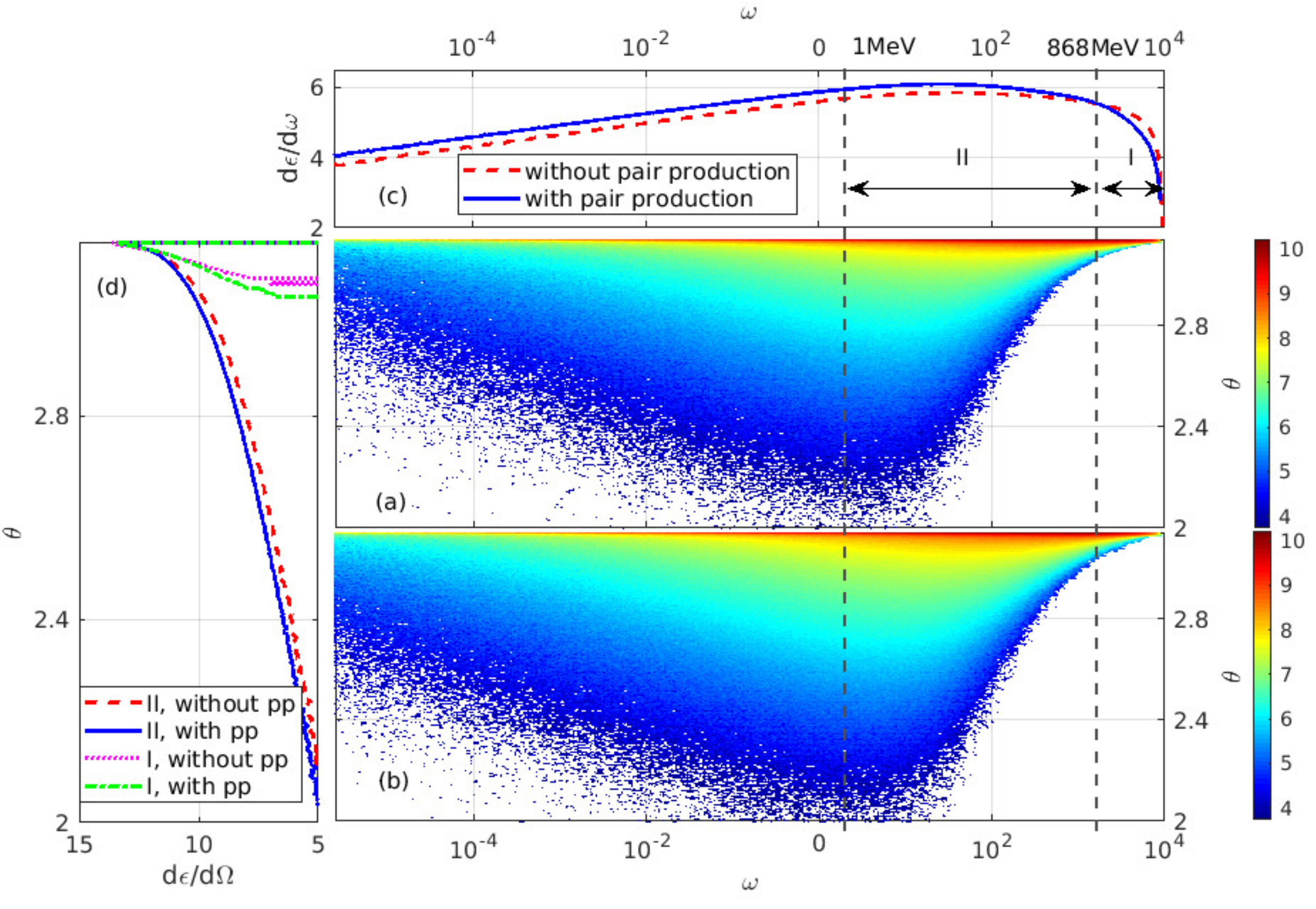}
		\end{center}
	\caption{Radiation energy distribution log$_{10}[d\varepsilon/d\omega/d\Omega]$ rad$^{-1}$ vs radiation photon energy $\omega$  and emission polar angle $\theta$ for circular LG$_{01}$ mode, without (a) and with pair production (b); (c) Radiation spectral distribution with (blue-solid) and without (red-dashed) pair-production; (d) Angular distribution of radiation in region I ($\omega> 868$ MeV) and II (1MeV $<\omega<$ 868 MeV); $\xi=120, \gamma=10^4$. The energies are in units of $mc^2$.}
	\label{Fig. 3}
\end{figure}
\begin{figure} [b]
	\begin{center}
	\includegraphics[width=0.5\textwidth]{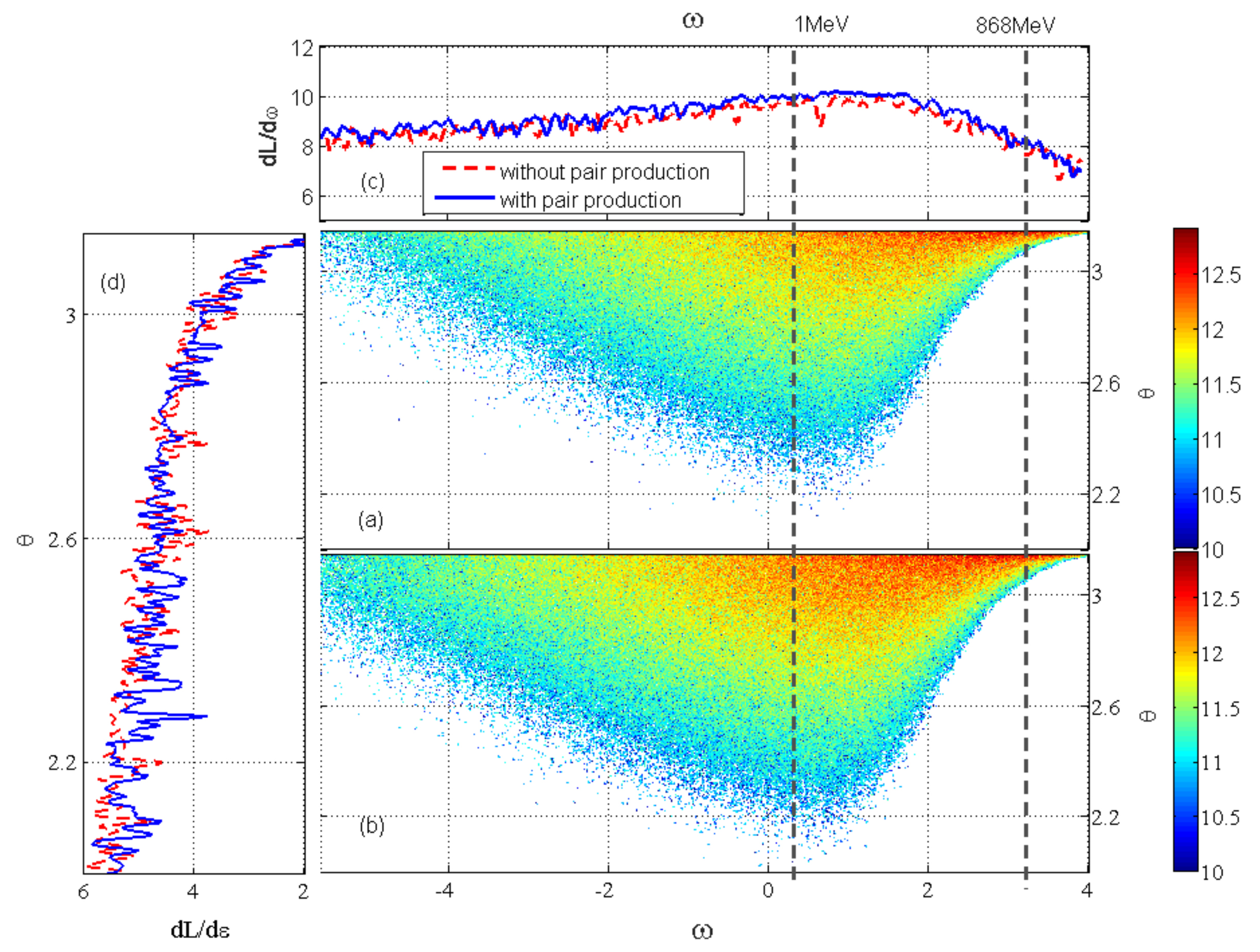}
	\end{center}
	\caption{Distribution of radiation OAM, log$_{10}[dL_\gamma/d\Omega/d\omega]$  vs  photon energy $\omega$ , and the emission polar angle $\theta$, for circular LG$_{01}$ mode, without (a) and with pair production (b); 
OAM vs photon energy (c), and  OAM per radiation energy $dL/d\varepsilon \equiv (dL_\gamma/d\Omega)/(d\varepsilon/d\Omega)$ vs $\theta$ (d),  with (solid blue lines) and without pair-production (red dashed lines). $\xi=120, \gamma=10^4$. OAM is in units of $\hbar$, and energies in units of $mc^2$.}
	\label{Fig. 4}
\end{figure}

We analyze the evolution of angular momentum described in Fig.~\ref{Fig. 2}. In all cases the angular momentum conservation is fulfilled to good accuracy  $(\ell+\sigma_z) n \hbar\approx  L_{e}+L_\gamma$, with the total OAM of the final electrons $L_e$, and that of the $\gamma$-photons $L_\gamma$. Here, the initial OAM of the electron beam vanishes; in the case with the pair production $L_e$ includes also the OAM of the created electrons and positrons. The total number of absorbed  laser photons due to the $\gamma$-photon emission is calculated via Eq. (7) of \cite{Supplement}, see the summary in Table \ref{Tab. 1}. In  QRDR the absorbed OAM of the laser photons is not fully transferred to the emitted photon beam but shared between the electrons and the emitted photons due to radiation reaction. The OAM share of  the $\gamma$-ray beam is $L_\gamma/(L_{e}+L_{\gamma})\sim 70\%$. This is in contrast to the idealized case of the  electron  interacting with a moderately strong laser field  \cite{katoh2017angular}, when $(\ell+\sigma_z) n \hbar=L_{\gamma}$ is fulfilled. The sign of radiation OAM is determined by  both  the laser and electrons' angular momenta, contrary to the case of linearly polarized LG at $\xi\sim 1$, where it is opposite to the  OAM absorbed laser photons \cite{petrillo2016compton}.

 \begin{figure}
	\begin{center}
	\includegraphics[width=0.5\textwidth]{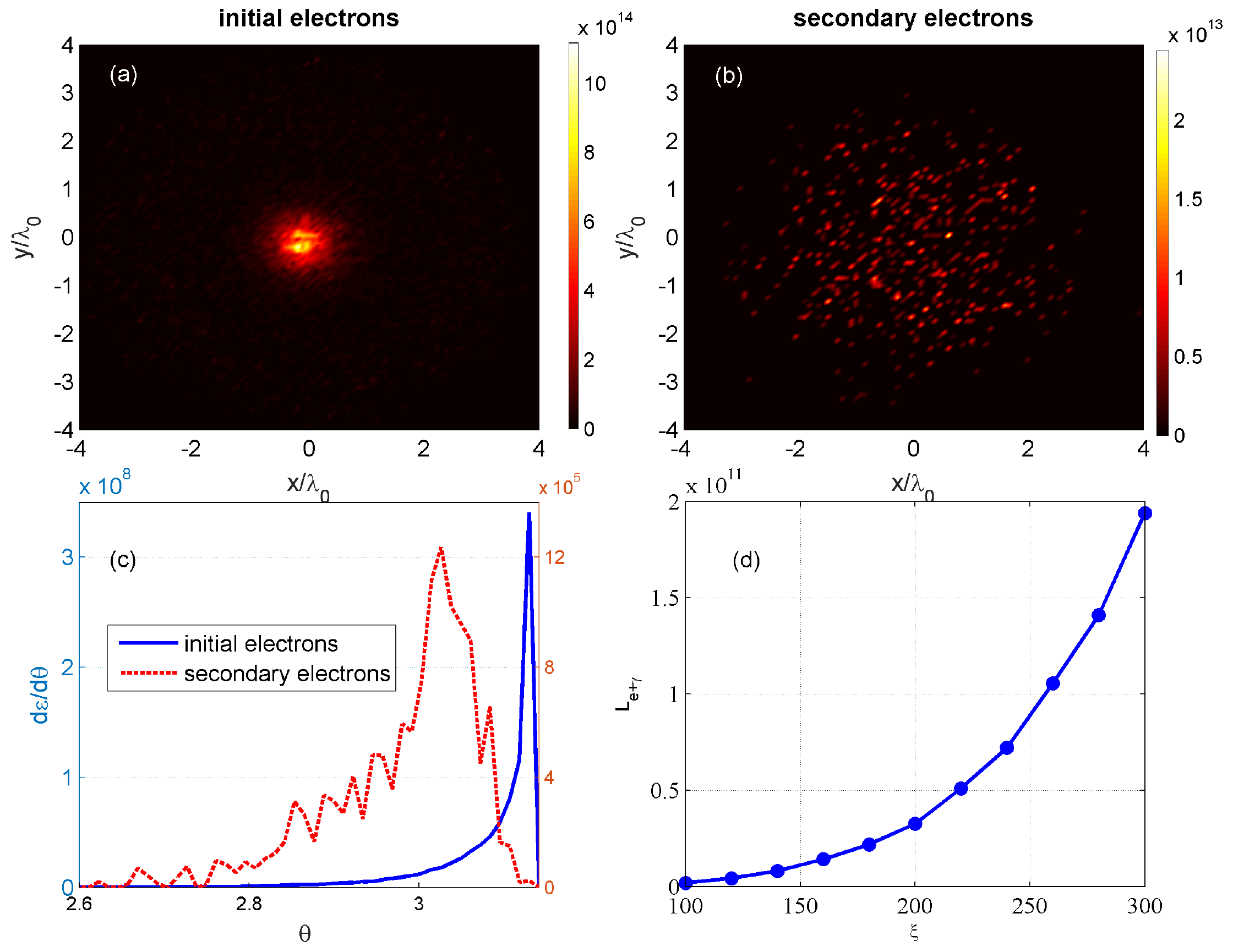}
	\end{center}
	\caption{Energy distribution log$_{10}[d\varepsilon/dx/dy]$cm$^{-2}$ in the transverse plane for the initial (a) and secondary (b) electrons at $t=10\tau_{p0}$;   (c) Energy of the initial (solid) and secondary (dashed) electrons in units of $mc^2$ 	vs polar angle. (d) Total OAM vs $\xi$, with $\chi=4$. 
		}
	\label{Fig. 5}
\end{figure}
 
One should underline that the transfer of OAM is determined by the total number of absorbed  laser photons during $\gamma$-photon emission, but not with  the total energy absorbed from the laser field during the interaction.  In fact, a part of the energy absorbed from the laser during photon emission is returned to the laser pulse after the turn off the laser field \cite{kibble1965frequency}. This ponderomotive energy transfer is not accompanied with an angular momentum transfer. The total energy absorbed from the laser field during the interaction in terms of the photon number $\tilde{n}$ can be evaluated from the energy-momentum conservation involving the electron 4-momenta before and after the interaction $p$ and $p'$, respectively: $p+\tilde{n} k=p'+k'$.   The energy difference corresponding to turn-on and turn-off is $\Delta n=n-\tilde{n}=\frac{\xi^2}{2}(\frac{1}{p'k}-\frac{1}{pk})$, which is responsible for upshifting of laser frequency \cite{kibble1965frequency}. As confirmed by Table \ref{Tab. 1}, the energy conservation is fulfilled by the total number of absorbed laser photons during the interaction $\tilde{n}$.  Here the OAM transfer is determined by  the  number of the absorbed laser photons $n$ due to $\gamma$-photon emission.

A comparison of the cases of different laser fields in Fig.~\ref{Fig. 2}  shows that a circular LG$_{01}$ mode is more favorable for generation of a $\gamma$-ray beam with a large OAM.
For a linear polarized LG$_{01}$ beam, the transverse electric field oscillates along $x$-direction, resulting in an oscillating OAM $L_z\approx-\sum_{i}y_ip_{xi}$, as shown in Fig. \ref{Fig. 2} (a). However, the final OAM for linear LG$_{01}$ mode is much less than that for circular polarization. This is because the absorbed photon number $n(t)\propto a(t)^2$ \cite{Supplement}, and circular polarization provides a more steady and larger absorption of twisted photons. Further, due to the spatial structure the LG$_{01}$ mode has three times larger energy than the Gaussian mode for same $\xi$, and $\hbar$ more OAM per photon, which results in larger OAM transfer, see Table \ref{Tab. 1}.

For the chosen parameters $\xi=120, \gamma=10^4$, the quantum parameter is rather large, $\chi\approx  4$, and the pair production effect is not negligible. Our results in Fig.~\ref{Fig. 2} demonstrate the counterintuitive role of pair production. Even though the energy of the $\gamma$-beam decreases due to pair production, the OAM of radiation shows an unexpected growth. For example, in the case of circular LG$_{01}$ the toal OAM of the $\gamma$-beam grows up from $9.46\times 10^{11} \hbar$ to  $2.02\times 10^{12} \hbar$, see Table~\ref{Tab. 1}.

To explain the changes of the OAM induced by pair production, energy and OAM spectra of circular LG$_{01}$ are shown in Figs. \ref{Fig. 3} and \ref{Fig. 4}, respectively.
Since high energy photons are depleted due to pair production, the total energy of photons with $\omega>868$ MeV decreases, as shown in Figs. \ref{Fig. 3} (c). Meanwhile, more low energy photons are generated due to secondary emission of pairs, leading to an increase of energy at $\omega<868$ MeV. Since the energy of pairs is smaller than that of the initial electrons, they oscillate with a larger amplitude  and radiate photons in a larger angle $\theta$, as shown in Fig. \ref{Fig. 5}. As the OAM per unit power is inversely proportional to $\theta$, as shown in Fig. \ref{Fig. 4} (d), the angular redistribution caused by pair production results in the increase of OAM for the total $\gamma$-beam. For photons with $\omega<868$ MeV, the increase of OAM can be seen clearly in Fig. \ref{Fig. 4} (c). However the increase of OAM is roughly counteracted by photon number depletion in $\omega>868$ MeV.

 Intense femtosecond vortex light with a few to hundred mJ has been experimentally generated at infrared wavelength \cite{sueda2004laguerre,lin2016generation}. Meanwhile, amplification of twisted laser intensities by 2-orders of magnitude is shown in plasma with stimulated Raman backscattering \cite{vieira2016amplification}. With these advanced techniques and posible further improvement such as multistaging, PW class twisted lasers used in our scheme can be realized in  near future. A well-collimated $\gamma$-photon beam can then be generated with a brightness of about $6.4\times 10^{22}$ photons/s/mm$^2$/mrad$^2$, an average OAM per photon of $2.7\times10^4\hbar$ and with an angular spread of $\Delta \theta \approx 0.05$ rad. Our scheme can produce gamma-ray beams with GeV photons, i.e. with photon energy  $\sim$ 10 times higher than the copropagating scheme \cite{liu2016generation} in a much smaller angular spread ($\sim$ 10 times smaller than the all-optical scheme \cite{gong2017ultra}).

We point out that the OAM of incoherent radiation in the ultrarelativistic QRDR is the property of the whole beam, rather than the property of single photons. If the initial beam contains $N_e=2\times10^{10}$ electrons, the total OAM of a $\gamma$-ray beam within $\theta>\pi-0.05$ rad is $L_\gamma \approx 1.9\times10^{16}\hbar$. The OAM of radiation can be largely increased by using a more intense laser field, see Fig.~\ref{Fig. 5}(d), however, at the expense of its collimation \cite{Supplement}.  The collective OAM of a $\gamma$-beam can have a significant mechanical impact. For instance, when this $\gamma$-rays are absorbed by a  disk of nuclear matter with radius of $2\lambda_0$, it would induce a rotation of the disk with a frequency of $\sim 100$ Hz \cite{Supplement}, similar to the frequency of a wave in the wave zone of the Crab pulsar.

Concluding, we showed a possibility for generation of well-collimated $\gamma$-ray beams with a large OAM in the ultrarelativistic quantum radiation-dominated regime, employing incoherent Compton scattering of twisted light by an electron beam. In contrast to the low intensity regime, each $\gamma$-photon is not in a certain OAM state, but, the total $\gamma$-ray beam carries a large angular momentum with respect to the beam axis. The OAM of the laser photons is transferred not only to $\gamma$-ray beam but also to electrons. Our results may have applications in simulating astronomical phenomena, and for interpretation of incoherent $\gamma$-ray generation around pulsars.

\bibliography{chen_paper} 

\end{document}